\documentclass[10pt]{iopart} 
\usepackage{iopams}

\usepackage{graphicx}
\usepackage{color}

\usepackage[normalem]{ulem} 

\newcommand{\ii}{\mathrm{i}} 

\newcommand{\tomega}{\tilde{\omega}}
\newcommand{\tphi}{\tilde{\phi}}

\newcommand{\vk}{ \bi{k} }

\newcommand{\imp}{ \mathrm{imp} }
\newcommand{\SC}{ \mathrm{SC} }
\newcommand{\kB}{ k_{\mathrm{B}} }
\newcommand{\Den}{ \mathrm{D} }

\begin{document}

\title[Thermodynamics of $s_{\pm}$-to-$s_{++}$ transition]{Thermodynamics of $s_{\pm}$-to-$s_{++}$ transition in iron pnictides in the vicinity of the Born limit}

\author{V.A. Shestakov$^1$, M.M. Korshunov$^{1}$}
\address{$^1$Kirensky Institute of Physics, Federal Research Center KSC SB RAS, 660036, Krasnoyarsk, Russia}
\ead{v\_shestakov@iph.krasn.ru}

\vspace{10pt}
\begin{indented}
\item[]\date{\today}
\end{indented}

\begin{abstract}
To study thermodynamical properties of the disorder-induced transition between $s_{\pm}$ and $s_{++}$ superconducting gap functions, we calculate the grand thermodynamic potential $\Omega$ in the normal and the superconducting states. Expression for the difference between the two, $\Delta\Omega$, is derived for a two-band model for Fe-based systems with nonmagnetic impurities. The disorder is considered in a $\mathcal{T}$-matrix approximation within the multiband Eliashberg theory. In the vicinity of the Born limit near the $s_{\pm}$-to-$s_{++}$ transition, we find two solutions obtained for opposite directions of the system's evolution with respect to the impurity scattering rate. By calculating the change in entropy $\Delta S$ and the change in electronic specific heat $\Delta C$ from $\Delta\Omega$, we show that such a hysteresis is not due to the time-reversal symmetry breaking state, but it rather points out to the first order phase transition induced by the nonmagnetic disorder. Based on the $\Delta\Omega$ calculations, phase diagram is plotted representing the energetically favourable $s_{\pm}$ and $s_{++}$ states and the transition between them. At finite temperature, a first order phase transition line there is limited by a critical end point. Above that point, the sharp $s_{\pm} \to s_{++}$ transition transforms to a crossover between $s_{\pm}$ and $s_{++}$ states.
\end{abstract}

\vspace{2pc}
\noindent{\it Keywords}: unconventional superconductors, iron pnictides, iron chalcogenides, impurity scattering, grand thermodynamic potential, first order phase transition



\ioptwocol

\section{Introduction}
Discovery of superconductivity in iron pnictides~\cite{y_kamihara_06, y_kamihara_08} boosted an interest in multiband systems~\cite{SadovskiiReview2008, IzyumovReview2008, IvanovskiiReview2008, JohnstonReview, StewartReview, HirschfeldKorshunov2011}. The ideas appeared some time ago in application for MgB$_2$~\cite{Golubov1995, Golubov1997} got a new life for iron-based superconductors~\cite{Mazin2008, ParkerKorshunov2008}. Order parameter with the $s_{\pm}$ gap function that changes its sign between different bands was suggested as the leading superconducting instability and got experimental support from neutron scattering~\cite{LumsdenReview, Korshunov2014eng, Inosov2016, Korshunov2018}, quasiparticle interference imaging~\cite{Hoffman2011, Hirschfeld2016}, and Andreev reflection measurements~\cite{KorshunovKuzmichev2022}.

A remarkable property giving prominence to iron based superconductors amongst a huge family of unconventional superconductors is their robustness to the suppression of superconductivity via nonmagnetic impurities~\cite{Karkin2009, Kim2010, Karkin2012, Reid2012, Cho2014, Prozorov2014, Strehlow2014, Li2016, KorshunovUFN2016, Surmach2017}. This property is associated with the possibility of changing structure of a superconducting order parameter upon the addition of nonmagnetic impurities~\cite{KorshunovUFN2016, EfremovKorshunov2011}. The change of the order parameter structure is seen as the transition from the $s_{\pm}$ state to the state with $s_{++}$ sign-preserving gap function.

The possibility of the experimental observation of this transition may be provided by the considering the London's penetration depth for a superconductor with nonmagnetic disorder~\cite{ShestakovKorshunovSUST2021}. There are at least two independent groups claiming the experimental observation of the transition~\cite{Schilling2016, Ghigo2018}.

The transition is significantly affected by the strength of an impurity potential~\cite{ShestakovKorshunovSUST2018, ShestakovKorshunovSymmetry2018}. Namely, for a weak scattering potential in the so-called Born limit, the transition is characterized by the abrupt change of the gap function while for stronger potentials it is smooth with one of the gaps going through zero. In the unitary limit of strong scattering impurity potential, the multiband analogue of the Anderson's theorem~\cite{p_anderson_1959, Golubov1995} is held, and the transition is absent. The nature of the abrupt transition near the Born limit remains unclear. To shed light on details of the $s_{\pm}$-to-$s_{++}$ transition in the vicinity of the Born limit, here we investigate thermodynamic properties of a two-band superconducting system. In particular, the grand thermodynamic potential (also known as Landau free energy) $\Omega$ and its temperature derivatives: entropy and specific heat. We show that the sharp change of the superconducting gap function is connected with the first order phase transition (PT) induced by the scattering of quasiparticles on the nonmagnetic impurity potential. A line of the first order phase transition in a phase diagram in axes (\textit{temperature}, \textit{impurity scattering}) has a critical end point (CEP) at temperature $T_\mathrm{CEP}$. Based on the analysis of results presented here and obtained earlier \cite{ShestakovKorshunovSUST2018, ShestakovKorshunovSymmetry2018}, CEP could be tuned into a quantum critical point (QCP) at zero temperature using a non-thermal parameter $\sigma$ (effective scattering cross section) connected with the strength of the scattering potential.

\section{Model and approach}
Here we use a two-band model of an iron-based superconductor with nonmagnetic impurities~\cite{KorshunovUFN2016, EfremovKorshunov2011} in terms of the $\xi$-integrated Green's functions $\hat{\mathbf{g}}(\ii\omega_n)$ defined within the combined band and Nambu spaces (which are denoted by \textbf{bold face font} and symbol `$\hat{\phantom{a}}$', respectively) and depending on the fermionic Matsubara frequency $\ii\omega_n$,
\begin{equation}\label{eq:GreensFunction}
 \hat{\mathbf{g}}(\ii\omega_n) = -\pi N_\alpha\frac{\ii\tomega_{\alpha n}\hat{\tau}_0 + \tphi_{\alpha n}\hat{\tau}_2}{\sqrt{\tomega_{\alpha n}^2 + \tphi_{\alpha n}^2}}\otimes\mathbf{1}_{\alpha\beta},
\end{equation}
where $\alpha = (a,b)$ is a band index, $N_\alpha$ is a density of states at the Fermi level in the normal state, $\ii\tomega_{\alpha n}$ and $\tphi_{\alpha n}$ are Matsubara frequency and superconducting order parameter, respectively, both renormalized by superconducting interaction and scattering on nonmagnetic impurities, $\hat{\tau}_{j}$ are the Pauli matrices within the Nambu space, and $\mathbf{1}_{\alpha\beta}$ is a unit matrix within the band space. Here and below, the following system of units is used: $\hbar = \kB = 1$. Thus, temperatures $T$ and frequencies $\omega_n = 2(n + 1)\pi T$ are given in units of energy. Note, that terms proportional to Pauli matrices $\hat{\tau}_1$ and  $\hat{\tau}_3$ are absent. The first one is omitted due to symmetry of the equations on the order parameter in the Nambu space \cite{FTT2024}, while the other one vanishes due to the $\xi$-integration procedure.

Matsubara frequencies and order parameter are self-consistently renormalized by the self-energy in the following manner,
\begin{eqnarray}
 &\ii\tomega_{\alpha n} = \ii\omega_n - \Sigma_{0}^{\SC}(\ii\tomega_{\alpha n}, \tphi_{\alpha n}) - \Sigma_{0}^{\imp}(\ii\tomega_{\alpha n}, \tphi_{\alpha n}),\label{eq:RenormOmega}\\
 &\tphi_{\alpha n} = \Sigma_{2}^{\SC}(\ii\tomega_{\alpha n}, \tphi_{\alpha n}) + \Sigma_{2}^{\imp}(\ii\tomega_{\alpha n}, \tphi_{\alpha n}),\label{eq:RenormPhi}
\end{eqnarray}
where part of the self-energy $\Sigma^{\SC}$ is connected with the superconducting interaction and depends on a $2\times2$ matrix of the coupling constants with elements $\lambda_{\alpha\beta}$ in the band space, while $\Sigma^{\imp}$ is related to nomnagnetic impurity scattering and calculated within the approximation of noncrossing diagrams --- the so-called $\mathcal{T}$-matrix approximation. The indices `0' and `2' refer to the corresponding Pauli matrices $\hat{\tau}_i$. Equations (\ref{eq:RenormOmega}) and (\ref{eq:RenormPhi}) represent nothing but the Eliashberg equations for multiband superconductor with nonmagnetic impurities.

\section{Landau free energy}
Most generally, the Landau free energy is given by the Luttinger-Ward expression for a multiband system~\cite{LW1960II, Luttinger1960} that is being generalized for the case of a superconductor with nonmagnetic impurities:
\begin{equation}\label{eq:LW}
 \eqalign{ \fl\Omega_{\mathrm{S}}(T) &= -T\sum_{\omega_n,\vk}\Tr\left[ \ln{\lbrace-\hat{\mathbf{G}}^{-1}(\vk,\ii\omega_n)\rbrace}\right. \cr &+ \left.\hat{\mathbf{\Sigma}}(\vk,\ii\omega_n)\hat{\mathbf{G}}(\vk,\ii\omega_n) \right] + \Omega'_\SC(T) + \Omega'_\imp(T),}
\end{equation}
where Green's function $\hat{\mathbf{G}}$ and self-energy $\hat{\mathbf{\Sigma}}$ have the general form and depend on momentum $\vk$ and Matsubara frequency $\omega_n$, $\Tr[...]$ is the trace over all subspaces (Nambu, band), sum over momenta $\sum_{\vk}$ denotes the integration over the whole first Brillouin zone, $\sum_{\vk}\leftrightarrow\int_{1\mathrm{BZ}}\mathrm{d}^3k/(2\pi)^3$. Here $\Omega'_\SC(T)$ and $\Omega'_\imp(T)$ are the `superconducting' and `impurity' parts of the Luttinger-Ward functional calculated within the same diagrammatic expansion as used in calculating $\ii\tomega_{\alpha n}$ and $\tphi_{\alpha n}$ in equations~(\ref{eq:RenormOmega}) and~(\ref{eq:RenormPhi}).
More precisely, $\Omega'_\SC(T)$ is determined by the interactions, electron-phonon or exchange of spin-fluctuations or both, which contribute to the formation of the superconducting state. And $\Omega'_\imp(T)$ arises due to the impurity scattering. Despite its name, $\Omega'_\SC(T)$  is not confined to the superconducting state since the effective interaction between quasiparticles present in the normal state and renormalizes Matsubara frequencies there.

For the normal state, free energy $\Omega_\mathrm{N}$ has a similar form. The only differences that all the quantities are calculated in the normal state ($\tphi_{\alpha n} = 0$). In practice, it is convenient to consider the difference between free energies of the superconducting and normal states,
\begin{equation}\label{eq:DeltaOmega}
 \Delta\Omega(T) = \Omega_{\mathrm{S}}(T)-\Omega_{\mathrm{N}}(T).
\end{equation}

Within the two-band model considered here, the difference has the form
\begin{equation}\label{eq:DeltaOmega2}
 \eqalign{ \fl\Delta\Omega(T) &= - \pi T\sum_{\omega_n}\sum_{\alpha = a,b} N_{\alpha}\left[ \frac{\omega_n\tomega_{\alpha n}}{\sqrt{\tomega_{\alpha n}^2 + \tphi_{2\alpha}^2(\ii\omega_n)}}\right. \cr &+ \left. \sqrt{\tomega_{\alpha n}^2 + \tphi_{2\alpha}^2(\ii\omega_n)} -\left|\omega_n\right| - \left|\tomega_{\alpha n}^{\mathrm{N}}\right| \right]
 + \Delta\Omega'(T),}
\end{equation}
where
\begin{equation}\label{eq:DeltaOmega2p}
 \eqalign{ \fl\Delta\Omega'(T) &= \pi T N_a\Gamma_a\sum_{\omega_n}\left[ \frac{2\sigma(1-\eta^2)^2 + (1-\sigma)\kappa_{\imp}}{2\Den_{\imp}}\right. \cr &- \left. \frac{2\sigma(1-\eta^2)^2 + (1-\sigma)\kappa_{\imp}^{\mathrm{N}}}{2\Den_{\imp}^{\mathrm{N}}} \right] \cr &- n_{\imp}T\sum_{\omega_n}\ln{\left( \frac{\Den_{\imp}}{\Den_{\imp}^\mathrm{N}} \right)},}
\end{equation}
\begin{equation}\label{eq:kappa.imp}
 \kappa_{\imp} = \eta^2\frac{N_a^2 + N_b^2}{N_aN_b} + 2\frac{\tomega_{an}\tomega_{bn} + \tphi_{2an}\tphi_{2bn}}{\sqrt{\tomega_{an}^2 + \tphi_{2an}^2}\sqrt{\tomega_{bn}^2 + \tphi_{2bn}^2}},
\end{equation}
\begin{equation}\label{eq:D_imp}
 \Den_{\imp} = (1-\sigma)^2 + \sigma^2(1-\eta^2)^2 + \sigma(1-\sigma)\kappa_{\imp}.
\end{equation}
$\kappa_{\imp}^{\mathrm{N}} = \Bigl.\kappa_{\imp}\Bigr|_{\tphi_{an} = \tphi_{bn} = 0}$, $\Den_{\imp}^{\mathrm{N}} = \Bigl.\Den_{\imp}\Bigr|_{\tphi_{an} = \tphi_{bn} = 0}$, $\eta = u/v$ is a ratio between interband ($u$) and intraband ($v$) components of the impurity potential, $\sigma$ is an effective scattering cross section,
\begin{equation}\label{eq:sigma}
 \sigma = \frac{\pi^2N_aN_bu^2}{1 + \pi^2N_aN_bu^2},
\end{equation}
$n_\imp$ is a concentration of impurities, and $\Gamma_a$ is an impurity scattering rate,
\begin{equation}\label{eq:Gamma_a}
 \Gamma_a = \frac{2n_{\imp}\sigma}{\pi N_a} = 2n_{\imp}\pi N_b u^2(1-\sigma),
\end{equation}
controlling the disorder in the system. The effective cross section represents strength of the scattering potential of impurities and varies from $0$ for weak scattering potential ($\pi u N_\alpha \ll 1$) in the Born limit to $1$ in the unitary limit of strong impurity scattering ($\pi u N_\alpha \gg 1$). In the Born limit, after taking the limit $\sigma \to 0$ under logarithm, the two last terms in~(\ref{eq:DeltaOmega2}) chancels out. Thus, $\Delta\Omega$ becomes only implicitly dependent on impurities through self-consistent solution of the equations~(\ref{eq:RenormOmega}) and~(\ref{eq:RenormPhi}).

\section{Results and Discussion}
In calculations below, we use the following values for the components of the coupling constant matrix $\lbrace \lambda_{aa}, \lambda_{ab}, \lambda_{ba}, \lambda_{bb} \rbrace = \lbrace 3.0, -0.2, -0.1, 0.5 \rbrace$. It gives the superconducting state below critical temperature in the clean limit $T_{c0} = 40$~K with the $s_{\pm}$ order parameter's structure and a positive coupling constant averaged over bands, $\langle \lambda \rangle = (N_a[\lambda_{aa} + \lambda_{ab}] + N_b[\lambda_{ba} + \lambda_{bb}]) / (N_a+N_b)$. It is the state in which scattering on nonmagnetic impurities leads to the $s_{\pm} \to s_{++}$ transition. We assume the impurity scattering to occur in the interband channel only, $\eta = 0$, since it was previously shown the nonzero intraband scattering potential has no influence on the superconducting state in the Born limit and only shifts the transition point to higher values of $\Gamma_a$~\cite{ShestakovKorshunovSUST2018}. The density of states is chosen to be $N_a = 1.0656$~eV$^{-1}$ and $N_b = 2 N_a$, so the total density of states $N = N_a + N_b$ is close to the one obtained within the first-principle calculations~\cite{Ferber2010, Yin2011, Sadovskii2012}.

\subsection{Hysteresis in solutions of equations for the superconducting order parameter}
\begin{figure}[h]
 \centering
 \includegraphics[width=0.5\textwidth]{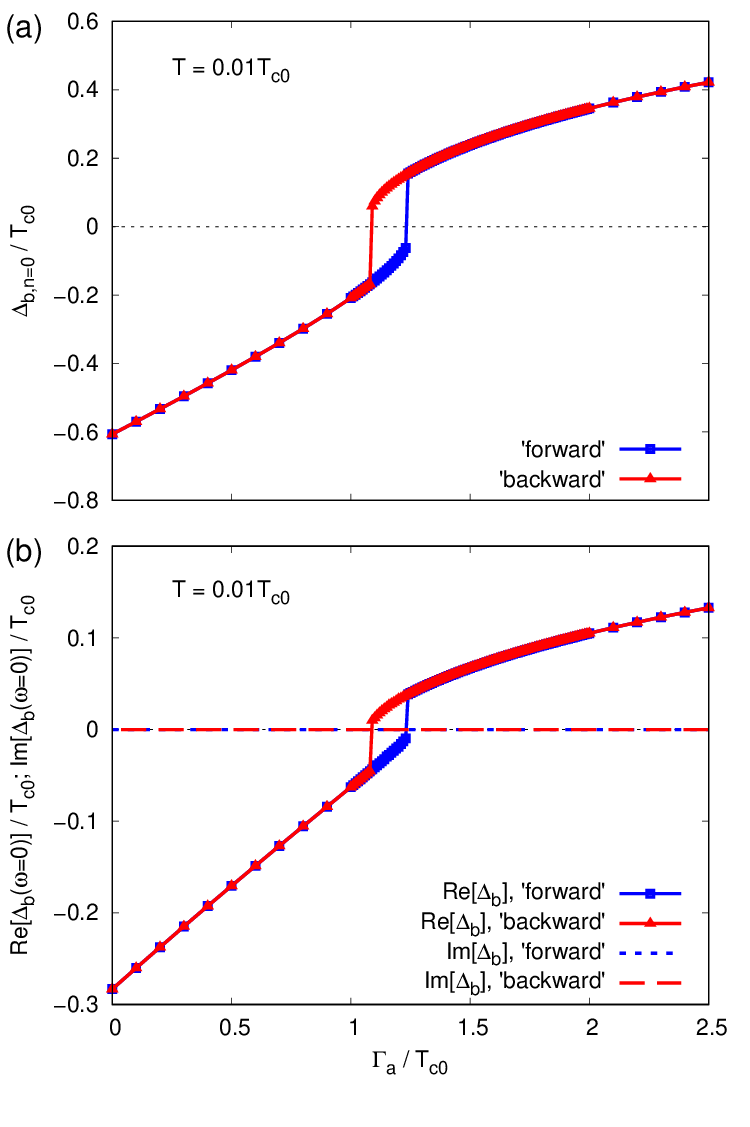}
 \caption{Dependence of the superconducting gap in the Born limit for the first Matsubara frequency $n = 0$ (a) and zero real frequency $\omega=0$ (b) in band $b$ on $\Gamma_a$. `Forward' denotes evolution of the system from the clean to the dirty limit, and `backward' denotes reverse evolution starting from the initially disordered state. Hereinafter, values of gap function $\Delta_b$ and the impurity scattering rate $\Gamma_a$ (as well as temperatures) are in units of the critical temperature for the clean superconductor $T_{c0}$.}
 \label{fig:hysteresisBorn}
\end{figure}

Earlier, we shown that the transition between $s_{\pm}$ and $s_{++}$ states for the effective cross section $\sigma < 0.12$ and temperatures $T < 0.1T_{c0}$ proceed in a discontinuous manner, i.~e. the order parameter within one of the bands changes its sign abruptly \cite{ShestakovKorshunovSUST2018}. Now we show that the Eliashberg equations~(\ref{eq:RenormOmega}) and~(\ref{eq:RenormPhi}) within the range $0 < \sigma < 0.12$ have two types of solutions
at $T < 0.1 T_{c0}$. They are obtained by moving in opposite directions along $\Gamma_a$ axis. To obtain the first type of solutions, we solve the Eliashberg equations in the clean limit, $\Gamma_a = 0$. Next, we add impurities, with solutions for the clean-limit being used as the initializing values. At the following step, we increase $\Gamma_a$ and use the previous solutions for impure system as the initializing values. Thus, we construct evolution of a superconductor from the clean to the disordered state. The second type of solutions is obtained by reversion of the impurity evolution direction, i.~e. we start from a dirty limit ($\Gamma_a = 6T_{c0}$) and `purify' the system up to the clean limit by decreasing $\Gamma_a$ thus using the solutions at higher $\Gamma_a$ as the initializing values for calculating solutions at lower $\Gamma_a$. These two types of solutions are illustrated in figure~\ref{fig:hysteresisBorn}(a), where the superconducting gap function $\Delta_{b,0}(\Gamma_a, T = 0.01 T_{c0})$ in band $b$ for the first Matsubara frequency ($n = 0$) is shown in the Born limit. 
We see a hysteresis effect between solutions for the system evolving from a clean state to a dirty state (denoted by `forward') and vice versa (denoted by `backward').

What is the origin of the observed behaviour? One of the possibilities is the existence of a time-reversal symmetry breaking (TRSB) superconducting state, such as $s + \ii s$, induced by nonmagnetic impurities~\cite{Garaud2017, Silaev2017, Garaud2018}. We can make the following suggestion: there is another, complex, solution for the superconducting gap right in the area of hysteresis. Since our calculations were performed on the Matsubara axis, we have to make analytical continuation to real frequencies. Pad\'{e} approximation for the gap function was applied and the resulting real and imaginary parts of the superconducting gap $\Delta_b$ as a functions of $\Gamma_a$ at zero real frequency $\omega$ are shown in figure~\ref{fig:hysteresisBorn}(b). Real part $\mathrm{Re}\Delta_b(\omega=0)$ exhibit the same hysteresis, while the imaginary part $\mathrm{Im}\Delta_b(\omega=0)$ vanishes for all values of $\Gamma_a$. Therefore, there is no TRSB-state here.

\begin{figure}[h]
 \centering
 \includegraphics[width=0.5\textwidth]{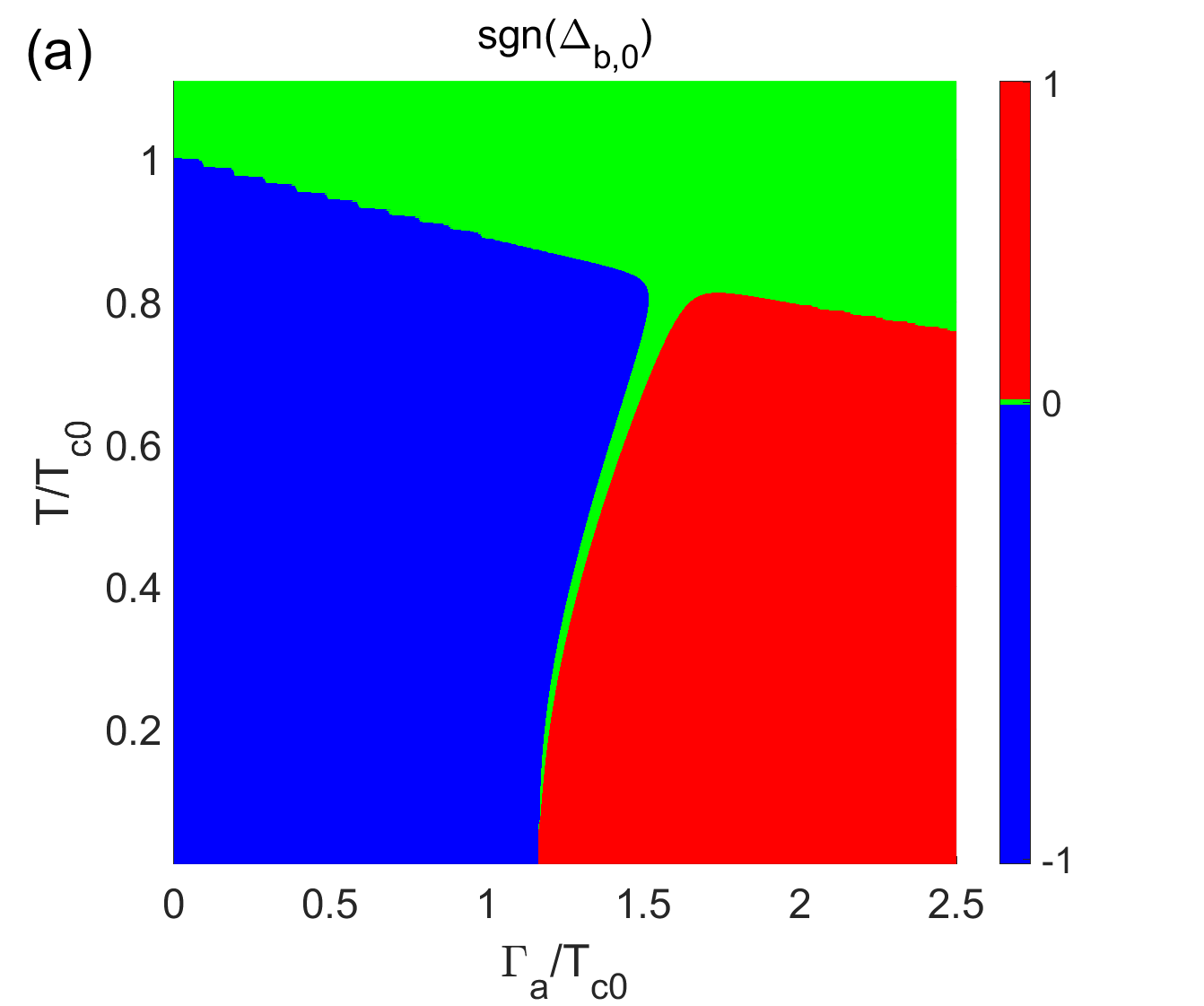}
 \includegraphics[width=0.5\textwidth]{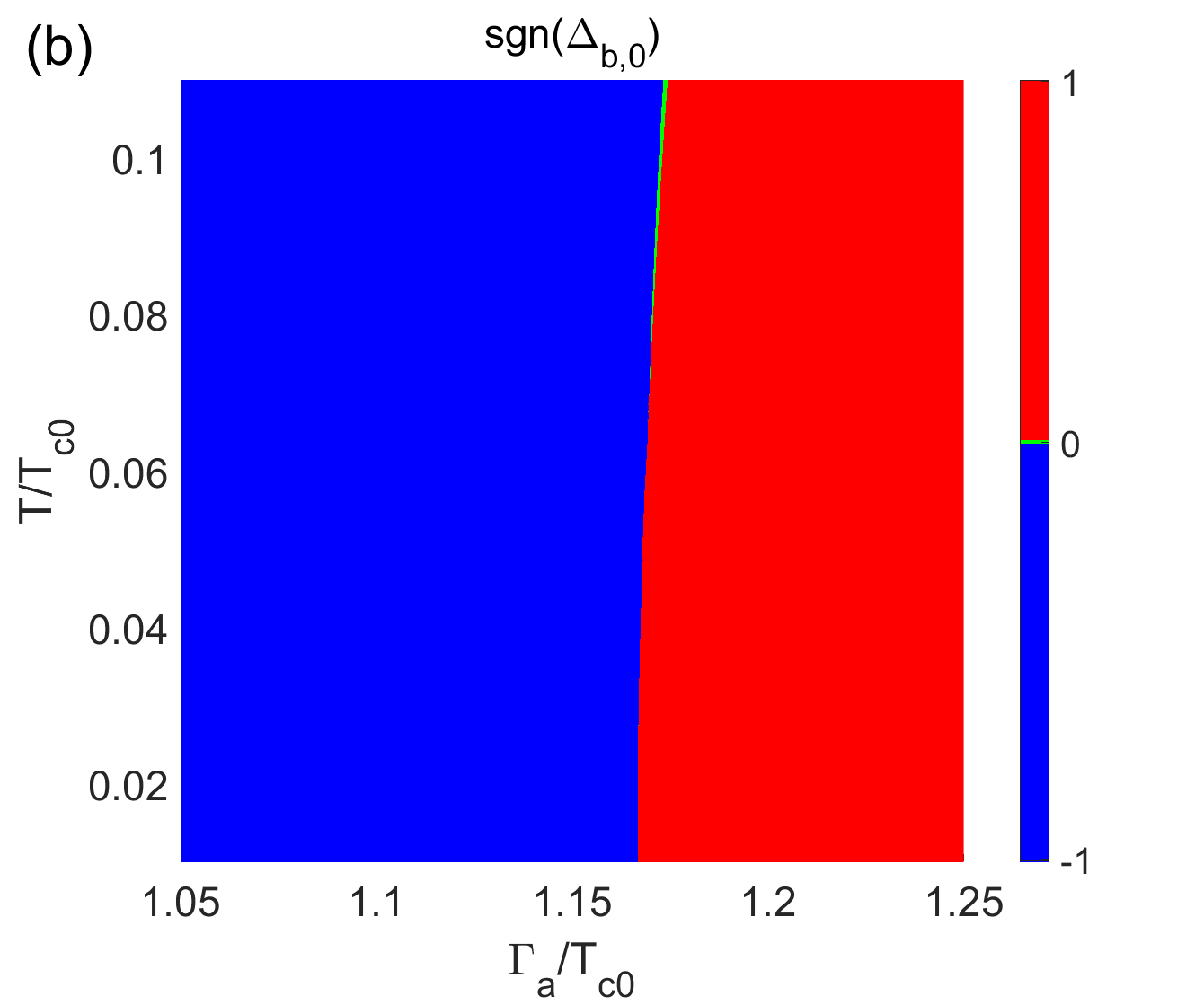}
 \caption{Phase diagram for the superconducting gap $\Delta_{b,0}$ in band $b$ for the first Matsubara frequency ($n = 0$) in axes ($T, \Gamma_a$). Colours represent sign of $\Delta_{b}$: red is for positive, blue is for negative, and green is for zero values. Panel (b) shows area close to the discontinuous $s_{\pm}$-to-$s_{++}$ transition. In panel (a), a step-like pattern of $T_c$ line (the boundary between green colour and others) is due to non-uniformity dense of computational grid.}
 \label{fig:phaseDiagramBorn}
\end{figure}

\begin{figure}[h]
 \centering
 \includegraphics[width=0.5\textwidth]{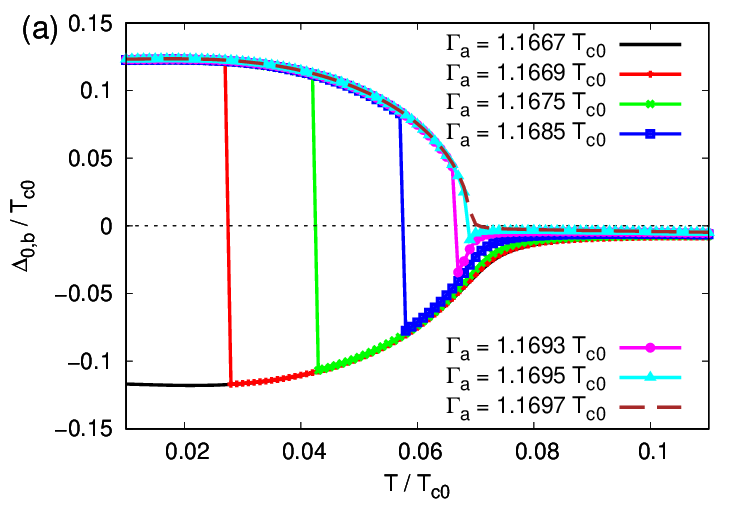}
 \includegraphics[width=0.5\textwidth]{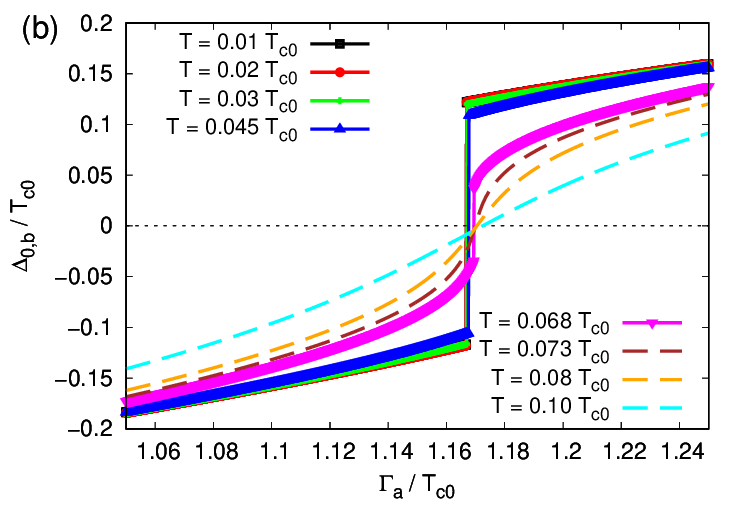}
 \caption{Plots of the gap function $\Delta_{b,0}$ dependence on temperature $T$~(a) and impurity scattering rate $\Gamma_a$~(b) corresponding to the phase diagram in figure~\ref{fig:phaseDiagramBorn}(b).}
 \label{fig:Deltab}
\end{figure}

Thus, what we observe, is the competition of two states represented by two solutions. Natural way to choose one of the solutions is to check which one is energetically favourable. Comparing the Landau free energies $\Delta\Omega$, we choose the solution with the lowest $\Delta\Omega$ from equation~(\ref{eq:DeltaOmega2}) and plot a phase diagram for the superconducting gap $\Delta_{b,0}$ in axes ($T,\Gamma_a$) in figure~\ref{fig:phaseDiagramBorn}.
Panels (a) and (b) represents different ranges of temperatures and $\Gamma$'s: $0.01 T_{c0} < T < 1.1 T_{c0}$,~$0 T_{c0} < \Gamma_a < 2.5 T_{c0}$ in the former and $0.01 T_{c0} < T < 0.11 T_{c0}$,~$1.05 T_{c0} < \Gamma_a < 1.25 T_{c0}$ in the later. The sign of the gap, which determines whether it is the $s_\pm$, $s_{++}$, or normal state, represented by color: blue is for the negative sign (which is opposite to the one for the second gap $\Delta_{a,0}$, $s_{\pm}$ state), red is for the positive sign (the same as for $\Delta_{a,0}$, $s_{++}$ state), and green is for values of $\Delta_{b,0}$ being close to zero. Note, since below $T_c$ the larger gap $\Delta_{a,0}$ is non-zero and positive, zero values of the smaller gap $\Delta_{b,0}(T<T_c)$ represents the so-called `gapless' superconducting state, not the normal state. In the phase diagram at low temperatures ($T < 0.07 T_{c0}$), we observe a vertical line separating the $s_{\pm}$ and $s_{++}$ states at $\Gamma_a \approx 1.16 T_{c0}$. Here $\Delta_{b,0}$ changes its sign abruptly. From figure~\ref{fig:phaseDiagramBorn}(b) we see that this line has a slope, which means that such an abrupt transition reveals itself also in the temperature dependence of $\Delta_{b,0}$. At temperatures $T > 0.07 T_{c0}$, the transition $s_{\pm} \to s_{++}$ is smooth and temperature dependent -- at these temperatures the line $\Delta_{b,0} = 0$ has even more prominent slope. In figure~\ref{fig:Deltab}, we show ``slices'' of the phase diagram from figure~\ref{fig:phaseDiagramBorn}(b) along the vertical and horizontal axes, i.~e. dependencies $\Delta_{b,0}(T)$ for a set of fixed $\Gamma$'s and $\Delta_{b,0}(\Gamma_a)$ for a set of fixed temperatures. Note the jumps in $\Delta_{b,0}(T)$ in panel (a), which indicate the discontinuous character of the $s_{\pm} \to s_{++}$ transition.

A similar result was obtained earlier~\cite{ShestakovKorshunovSymmetry2018} only for evolution `forward' in changing of nonmagnetic disorder. The main difference in the behaviour at low temperatures is that the phase diagram in~\cite{ShestakovKorshunovSymmetry2018} has a more prominent slope of the line for the abrupt sign changing of the superconducting gap function, since there are energetically unfavourable solutions are presented there. Here we avoided that by comparing the thermodynamic potentials of two possible solutions and choosing the one with the lowest energy.

\subsection{The grand thermodynamic potential $\Delta\Omega$, entropy $\Delta S$ and specific heat $\Delta C$ in the Born limit}

\begin{figure}[h]
 \centering
 \includegraphics[width=0.5\textwidth]{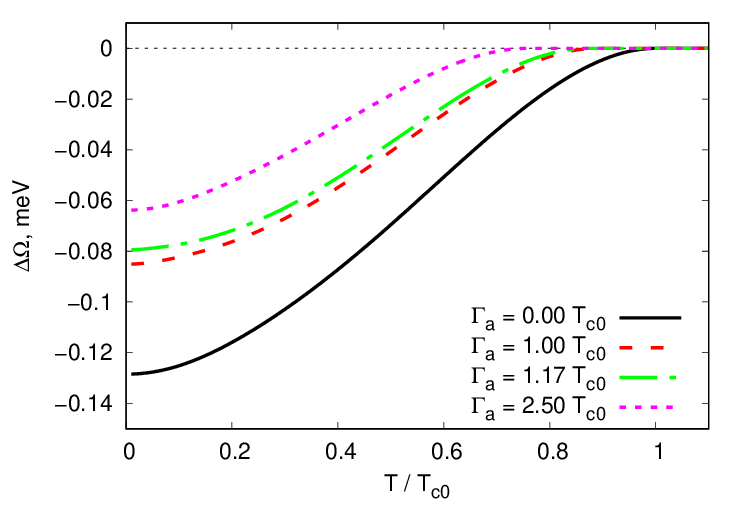}
 \caption{Temperature dependence of the Landau free energy $\Delta\Omega(T)$ for different values of $\Gamma_a$. There is no prominent signatures of the $s_{\pm} \to s_{++}$ transition.}
 \label{fig:DeltaOmegaSurfT}
\end{figure}

\begin{figure}[h]
 \centering
 \includegraphics[width=0.5\textwidth]{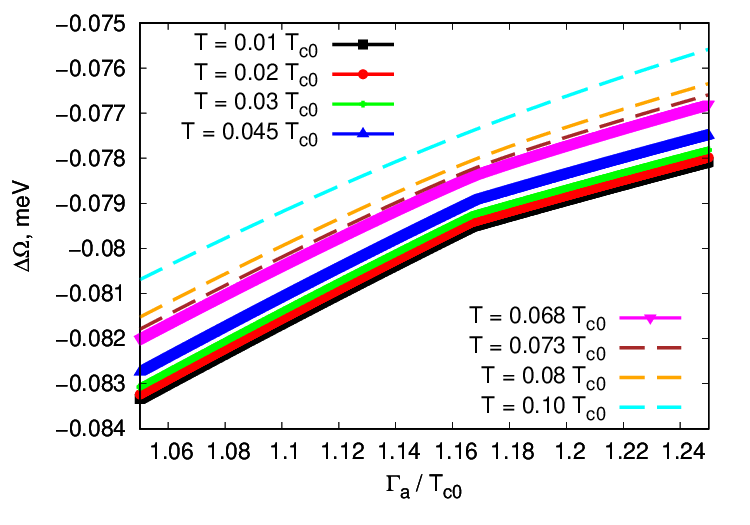}
 \caption{Dependence of $\Delta\Omega$ on $\Gamma_a$ for different temperatures corresponding to figure~\ref{fig:Deltab}(b). Although in figure~\ref{fig:Deltab}(b) there is a jump between negative and positive values of $\Delta_{b,0}$ at $T = 0.068T_{c0}$, we don't see a kink in $\Delta\Omega$ at this temperature.}
 \label{fig:DeltaOmega2d}
\end{figure}

Temperature dependence of the Landau free energy for different values of the impurity scattering rate $\Gamma_a$ is shown in figure~\ref{fig:DeltaOmegaSurfT}. It has a minimum at the lowest temperature $T = 0.01T_{c0}$ in the clean limit. Increasing temperature or $\Gamma_a$ gradually increases the free energy until the difference $\Delta\Omega(T)$ become equal to zero in the normal state. As seen in figure~\ref{fig:DeltaOmegaSurfT}, there is no prominent signatures of the $s_{\pm} \to s_{++}$ transition in the temperature dependence of $\Delta\Omega$. Nevertheless, the abrupt transition between $s_{\pm}$ and $s_{++}$ states reveals itself in the Landau free energy as a kink in dependence of $\Delta\Omega$ on $\Gamma_a$, see figure~\ref{fig:DeltaOmega2d}. The smooth $s_{\pm} \to s_{++}$ transition at $T > 0.1T_{c0}$ does not manifest itself directly in $\Delta\Omega$.

Such a peculiarity in dependence of the grand thermodynamic potential on non-thermal parameter $\Gamma_a$, which is accompanied by the hysteresis in solutions of the Eliashberg equations for the superconducting gap functions, points out to the first order phase transition induced by a nonmagnetic disorder.

\begin{figure}[h]
 \centering
 \includegraphics[width=0.5\textwidth]{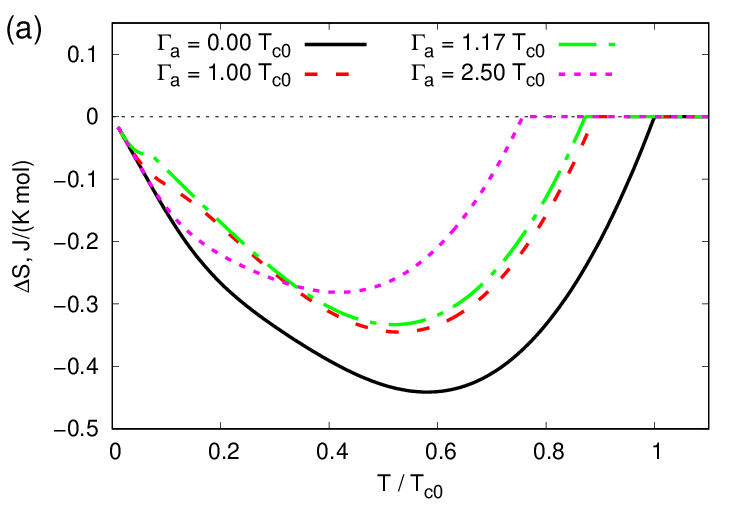}
 \includegraphics[width=0.5\textwidth]{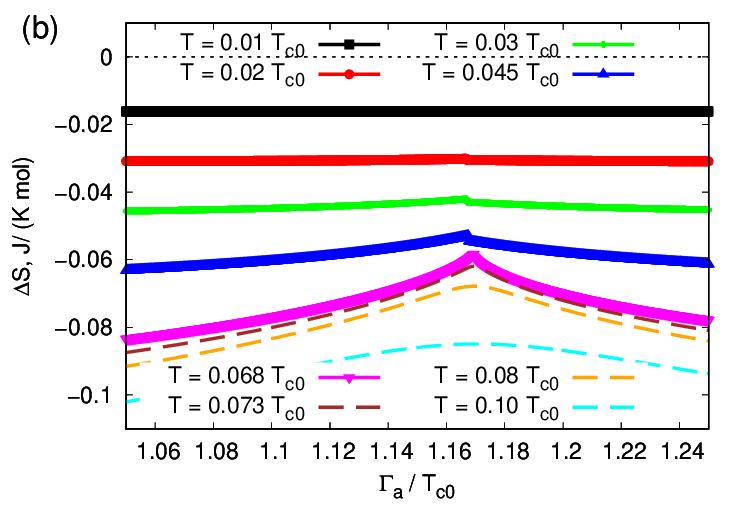}
 \caption{Entropy $\Delta S$ dependence on temperature $T$~(a) and impurity scattering rate $\Gamma_a$ at low temperatures $T \leq 0.1 T_{c0}$~(b). In panel (b), line for $T = 0.01 T_{c0}$ has a discontinuity that is not seen at the scale presented.}
 \label{fig:DeltaS2d}
\end{figure}

With the Landau free energy being calculated, we can perform a numerical differentiation of it and calculate the change in entropy $S(T)$ upon the transition to the superconducting state,
\begin{equation}\label{eq:Entropy}
 \Delta S(T) = - \frac{\partial\Delta\Omega(T)}{\partial T},
\end{equation}
and the change in electronic specific heat $C(T)$,
\begin{equation}\label{eq:SpecificHeat}
 \Delta C(T) = - T \frac{\partial^2\Delta\Omega(T)}{\partial T^2}.
\end{equation}
Dependence of entropy on temperature and impurity scattering rate $\Gamma_a$ is shown in figure~\ref{fig:DeltaS2d}(a) and (b), respectively. As it can be seen, the kink in $\Delta\Omega$ leads to the peak in $\Delta S$ in dependencies $\Delta S(T)$ for $\Gamma_a=\mathrm{const}$ and $\Delta S(\Gamma_a)$ for $T=\mathrm{const}$.

\begin{figure}[h]
 \centering
 \includegraphics[width=0.5\textwidth]{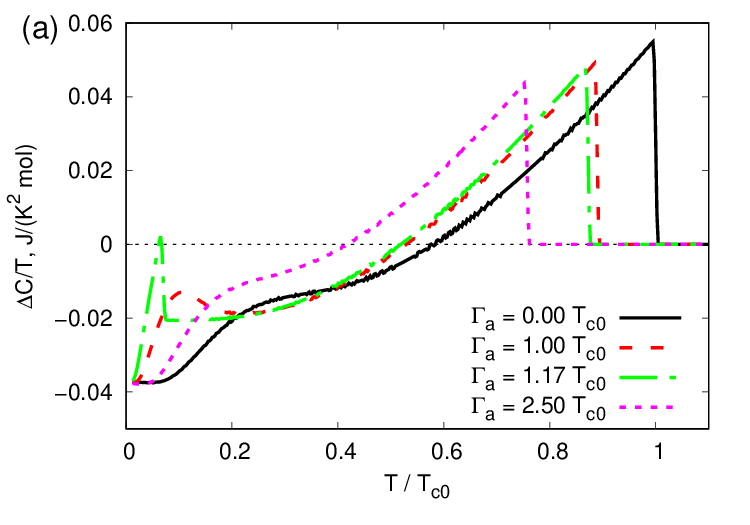}
 \includegraphics[width=0.5\textwidth]{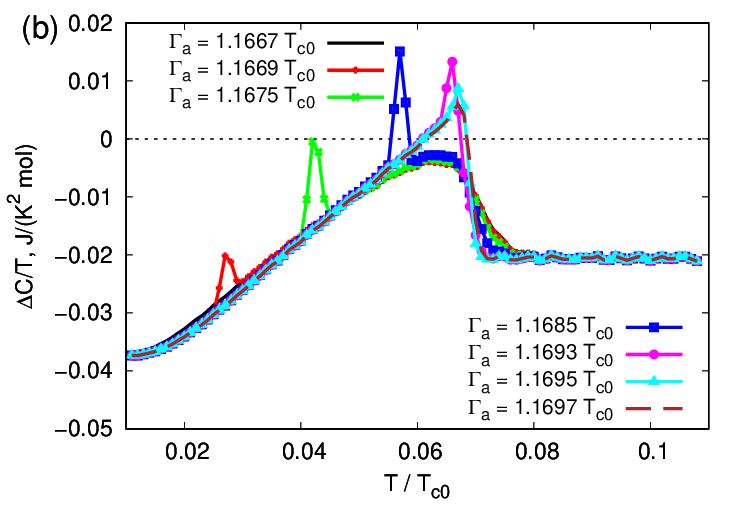}
 \includegraphics[width=0.5\textwidth]{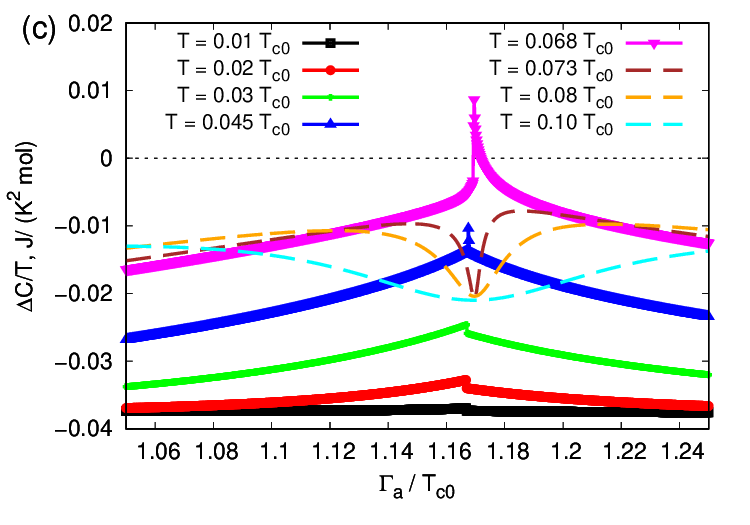}
 \caption{Electronic specific heat $\Delta C/T$ temperature dependence in the wide~(a) and in the narrow~(b) ranges of $T$, and its dependence on impurity scattering rate $\Gamma_a$ at low temperatures $T \leq 0.1 T_{c0}$~(c). Panel (b) demonstrates details of the low temperature behaviour of specific heat, where two type of peaks are seen: (i) those from discontinuity in sign changing of $\Delta_{b}$ emerging at different temperatures for different $\Gamma_a$, and (ii) peaks from fast decreasing of $\Delta_b$ magnitude at almost the same temperature $T\approx 0.07T_{c0}$ for different $\Gamma_a$.}
 \label{fig:DeltaC2d}
\end{figure}

Electronic specific heat $\Delta C / T$ is shown in figure~\ref{fig:DeltaC2d}. Its temperature dependence in panel (a) has two peculiarities. The first one is at $T_c$ for all values of $\Gamma_a$ and it is due to the second order phase transition from normal to superconducting state. The second one is at low temperatures $T < 0.07 T_{c0}$ for $1.1669 T_{c0} < \Gamma_a < 1.1695 T_{c0}$, see panel (b), and it is connected with the first order phase transition. Here we see a series of peaks for different values of $\Gamma_a$, which are due to the discontinuity in changing the sign of the gap function $\Delta_{b}$ across the first order PT. Note that the peaks are at the same temperatures, as the jumps in $\Delta_{b}$ shown in figure~\ref{fig:Deltab}(a). This series ends with a single peak at $T \approx 0.68 T_{c0}$ for all $\Gamma_a$ values within range $1.1669 T_{c0} < \Gamma_a < 1.1695 T_{c0}$. It is associated with changing the character of the $s_{\pm}$-to-$s_{++}$ transition from abrupt to smooth. That is, it's a critical end point $T_\mathrm{CEP}$, in which the low temperature first order PT transforms to a $s_{\pm} \to s_{++}$ crossover at higher temperatures.

In figure~\ref{fig:DeltaC2d}(c), we see the evolution of the low temperature features in $\Delta C(\Gamma_a) / T$ with temperature approaching $T_\mathrm{CEP}$ and then crossing it. A small discontinuity at low temperatures, $T < 0.04T_{c0}$, gradually becomes a divergent peak at the critical end point. Above $T_\mathrm{CEP}$, the peak splits and a pair of new smooth peaks, being gradually smeared, eventually disappears at $T > 0.5 T_{c0}$.

\section{Conclusions}
For multiband superconductor with nonmagnetic impurities in a close vicinity of the abrupt transition between $s_{\pm}$ and $s_{++}$ states, Eliashberg equations can have a set of non-unique solutions near the Born limit ($\sigma < 0.12$). Two different solutions are obtained for opposite directions of the system's evolution with respect to the impurity scattering rate. Such a hysteresis exists within certain range of temperatures and impurity scattering rates. The hysteresis appears in the gap function at imaginary Matsubara frequencies ($\Delta_{b,0}$), as well as in the real part of the gap function at zero real frequency, $\mathrm{Re}\Delta_b(\omega=0)$. On the contrary, imaginary part $\mathrm{Im}\Delta_b(\omega=0)$ vanishes that indicates absence of the TRSB state proposed earlier in Refs.~\cite{Garaud2017, Silaev2017, Garaud2018}.
Since our results are obtained within the two band model and it was claimed that the TRSB state could be obtained within a three band model~\cite{Grinenko2020}, further studies of the TRSB state possibility are required in the multiband systems.

To disentangle ambiguity in solutions, we calculated the Landau free energy $\Delta\Omega = \Omega_\mathrm{S} - \Omega_\mathrm{N}$ and, by choosing solutions with the lowest energy, plotted the phase diagram of $\mathrm{sign}[\Delta_{b,0}(T,\Gamma_a)]$ representing the $s_{\pm}$ and $s_{++}$ states and the transition between them. There is almost a straight line of the abrupt $s_{\pm} \to s_{++}$ transition along the temperature axis going from $T_{\mathrm{min}} = 0.01T_{c0}$ to $T \approx 0.07 T_{c0}$. The abrupt transition is characterized by the discontinuous jump of the smaller gap while it crosses zero. At temperatures above the mentioned line, the sharp $s_{\pm}$-to-$s_{++}$ transition is changed to the smooth one with the smaller gap showing continuous evolution with the impurity scattering rate~\cite{ShestakovKorshunovSymmetry2018}. This result refines the previously obtained phase diagram~\cite{ShestakovKorshunovSymmetry2018} where calculations were performed only for the increasing amount of disorder in the system.

The Landau free energy itself does not exhibit any pronounced signs of the $s_{\pm}$-to-$s_{++}$ transition in its temperature dependence but has a kink in its dependence on the impurity scattering rate, i.~e. on the non-thermal parameter. Based on this and on the hysteresis in solutions of Eliashberg equations for the gap function, we conclude that the first order phase transition induced by nonmagnetic disorder takes place at low temperatures. Critical end point for the first order PT line on the phase diagram~\ref{fig:phaseDiagramBorn} is determined by the peak in the specific heat at temperature $T_\mathrm{CEP} \approx 0.68 T_{c0}$. Above $T_\mathrm{CEP}$, the smaller gap $\Delta_b$ changes its sign smoothly, i.~e. the first order PT transforms to the $s_{\pm} \to s_{++}$ crossover.

Earlier we have shown that the abrupt transition between $s_{\pm}$ and $s_{++}$ states become smooth with increasing effective cross section $\sigma$ above $\approx 0.12$ at fixed temperature $T = 0.01T_{c0}$~\cite{ShestakovKorshunovSUST2018}. Combining that fact with our current results, we assume the critical end point may be shifted to zero temperature by tuning the parameter $\sigma$. Thus it would become a quantum critical point. Such a mechanism is similar to the one for a metamagnetic transition tuned by a non-thermal parameter in the bilayer strontium ruthenate Sr$_3$Ru$_2$O$_7$~\cite{Grigera2001} and rare-earth titanates~\cite{Wang2022}. To answer the question on the existence of the quantum critical point, however, a further separate studies in the zero-temperature limit are necessary.

\ack
We would like to thank O.V. Dolgov and A.A. Golubov for many useful discussions. This work was carried within the state assignment of Kirensky Institute of Physics.

\section*{References}

\bibliographystyle{iopart-num}
\bibliography{tdpotfebs_2col_arxiv}

\end{document}